# ON THE ORIGIN OF THE METAL-INSULATOR TRANSITION IN 2D

V.M. Pudalov, *Institute for High Pressure Physics, Troitsk, 142092 Russia*
pudalov@ns.hppi.troitsk.ru

Two phenomena were recently observed in high-mobility Si MOS structures: (1) strong enhancement of the metallic conduction [1] at low temperatures, $T < 2K$, and (2) the scaling behavior of the temperature and electric field dependences of the resistivity [2]. These results evidence for the true metal-insulator transition in 2d, in apparent disagreement with the scaling theory [3]. Here we present a model that explains both effects in the framework of the spin-orbit interaction and provides a quantitative agreement with the experimental data.

Fig. 1 shows a set of the $\rho(T)$ curves typical for the high mobility samples at different electron densities $n$. In the range $T \geq 2K$, $\rho$ increases slowly with decreasing temperature, that is consistent with weak localization [3]. But as temperature is further decreased, $\rho$ sharply drops for the densities above a ``critical value'' $n_c(T)$. Thus the enhancement of metallic properties, evidently, overpowers the onset of localization visible at higher temperatures. No further evidence for localization is seen at temperatures down to 20mK. For the curves at lower densities, $n < n_c$, the resistivity grows continuously with decreasing temperature showing permanently localized state.

The low-temperature variation of $\rho(T)$ becomes less pronounced with decreasing sample mobility. Table 1 shows the peak mobility, maximal value of the resistance drop, $\rho(T=2K)/\rho(0)$, and other relevant parameters for different samples.

Table 1 Relevant parameters for different samples.

| Sample | Si-15 | Si-5 | Si-12a | Si-12b | Si-14 | Si-39 | Si-33 |
|---|---|---|---|---|---|---|---|
| $\mu$ ($10^3$ cm$^2$/Vs) | 71 | 36 | 33 | 30 | 19 | 5.1 | 1.4 |
| $\rho(2K)/\rho(0)$ | 11 | 6.5 | 6.5 | 6 | 2 | 1.1 | ~1 |
| $\Delta_s/(h/\tau_q)$ | $\geq 2$ | ~2.0 | ~1.9 | $\approx 1.8$ | | ~1 | |

Due to the broken reversal symmetry in the triangular potential well of the Si-MOS structure, the zero field spin degeneracy of 2d electrons is lifted and the energy spectrum consists of two branches [4]:

$$E = \frac{\hbar^2 k^2}{2m^*} \pm \alpha \cdot k$$

As seen from Table 1, in the high mobility samples the level broadening, $h/\tau_q$, becomes less than the corresponding energy splitting between the two branches $\Delta_s$. In these samples the dominant scattering process is the scattering with conservation of the spin direction in the real space. The letter presumes spin-flip with respect to the momentum direction and, hence, is the transition between (+) and (-) branches of the energy spectrum. This transition occurs across the spin-orbit gap; the corresponding exponential factor $exp(-\Delta/kT)$ reduces the scattering probability at low temperatures. The scattering probability was calculated by taking into account the energy and momentum conservation laws, the acoustic phonon spectrum, and the two measured parameters: (i) the spin splitting $\Delta_s = 3.6$ K in the limit of $n=0$, $H=0$ [5], and (ii) the spin level broadening, $h/\tau_q$, extracted from the Shubnikov-de Haas effect at low density.

The resulting calculated dependencies $\rho(T)$ are shown in Fig. 1 by solid lines. In this approach the metal-insulator transition in 2d-system is entirely due to the spin-orbit interaction, which modifies the system symmetry from orthogonal to symplectic [6] and suppresses the quantum interference contribution. As a result, the scaling β-function shifts upward and changes the sign at a critical point, thus causing the metal-insulator transition [7] at $H=0$. The letter may occur only when the spin-splitting is resolved, $\Delta_s > h/\tau_q$, i.e. in the systems with strong spin-orbit interaction and small level broadening.

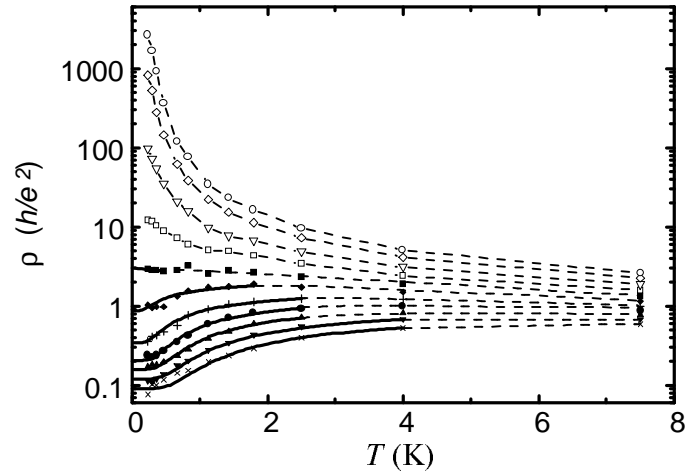

Fig.1. Typical dependencies $\rho(T)$ for high-mobility Si-MOS structure [2] at different electron density in the range 13.69 to $9.53 \times 10^{10}$ cm$^{-2}$.

In summary, the metal-insulator transition recently observed in high mobility Si-MOS structures, we believe, is the first experimental manifestation of the spin-orbit interaction induced transition in two dimensions.